\documentclass[a4paper,floatfix,rmp,twocolumn,showkeys,superscriptaddress]{revtex4}
\usepackage{graphicx}
\usepackage{url}
\usepackage{color}
\usepackage{float}

\begin{document}



\title{Synthetic associative learning in engineered multicellular consortia}

\providecommand{\ICREA}{ICREA-Complex Systems  Lab, Universitat Pompeu
  Fabra,   Dr    Aiguader   88,   08003   Barcelona}
  \providecommand{\IBE}{Institut de Biologia Evolutiva, CSIC-UPF, 
  Passeig Maritim de la Barceloneta, 37, 08003 Barcelona}
\providecommand{\SFI}{Santa Fe  Institute, 1399 Hyde  Park Road, Santa Fe NM 87501, USA}

\author{Javier Macia} \affiliation{\ICREA}    \affiliation{\IBE}
\author{Blai Vidiella} \affiliation{\ICREA}    \affiliation{\IBE}
 \author{Ricard V.   Sol\'e\footnote{Corresponding author}}   \affiliation{\ICREA} \affiliation{\IBE} \affiliation{\SFI}
  
\vspace{0.4 cm}
\begin{abstract}
\vspace{0.2 cm} 
Associative learning is one of the key mechanisms displayed by living organisms in order to adapt to 
their changing environments. It was early recognized to be a general trait of complex multicellular organisms 
but also found in "simpler" ones. It has also been explored within synthetic biology 
using molecular circuits that are directly inspired in neural network models 
of conditioning. These designs involve complex wiring diagrams to be implemented within 
one single cell and the presence of diverse molecular wires become a challenge that might be 
very difficult to overcome. Here we present three alternative circuit designs based on two-cell 
microbial consortia able to properly display associative learning responses 
to two classes of stimuli and displaying long and short-term memory (i. e. the association can be lost 
with time). These designs might be a helpful approach for 
engineering the human gut microbiome or even synthetic organoids, defining a new class of decision-making 
biological circuits capable of memory and adaptation to changing conditions. The potential implications 
and extensions are outlined.
\end{abstract}


\maketitle


\section{Introduction}


A specially important component of adaptation in nature is based on the capacity of some living beings to 
respond to external signals by a proper combination of repeated exposure to 
stimuli and the potential for storing memories.  One classical example is provided 
by early experiments on conditioning, also known as {\em associative learning} (AL) 
and is one particularly important example of a general class of processes involving 
{\em associative memory} (Walters et al 1979; Hassoun 1993). 
In these experiments, a given animal is known to respond automatically 
to an unconditioned stimulus (US) such as air puff in the eye that leads to eyelid closure. 
Instead, another stimulus such as a weak noise is unlikely to elicit a response. This would be an 
example of a conditioned stimulus (CS). In a nutshell, associative learning occurs when 
both stimuli are simultaneously presented, in such a way that a repeated exposure to 
both stimuli creates a cognitive link.  At some point 
the exposure to only CS leads to the response that was originally limited to US: the weak noise 
triggers eyelid closure. 

Conditional learning is part of the enormous potential exhibited by organisms 
having neuronal systems and might have been a crucial innovation in the 
evolutionary history of multicellularity (Ginsburg and Jablonka 2010). Many forms of adaptation are grounded in neuronal 
circuits capable of creating correlations between different events, providing a plastic 
and reliable way of predicting future changes (Grossberg 1988; 
Gerstner and Kistler 2002). Most of these examples involve 
the presence of a neural circuitry, but the phenomenon seems to be also at work 
in non-neural agents. For example, microorganisms are capable of dealing with environmental correlates 
and perform decision making tasks (Ben-Jacob 2004 ; Tagkopoulos et al 2008; Ben-Jacob 2009; Mitchell et al 2009; Reid et al 2015). This includes in particular molecular mechanisms 
responsible for information processing (Bray 1995; Buchler et al. 2003). A relevant question here is how could we synthetically enhance the cognitive complexity of microorganims and how this can gives insight into the origins and evolution of 
microbial inteligence (Sol\'e 2016). The potential for designing living systems has been rapidly improved in the last decade. Among the most promising areas where such engineering of microbial intelligence can be crucial is the engineering of the human microbiome (Huttenhower et al 2012;  Ackerman 2012; Ruder et al 2011). Treatments and recoveries from disturbance have been shown to be transitions among alternative states (Costello 2012, Pepper and Rosenfeld 2012). Mounting evidence 
reveals that this complex ecosystem is relevant in many pathological states and that 
engineered microbes could be designed to detect and cure microbiome-related disorder (Sonnenburg 2015). 
Since we are often dealing with complex diseases, such as inflammatory processes, these need to be 
smart bacteria, capable of delivering drugs under the required conditions but also shut off once the 
inflammation is eliminated. This is more obvious if we take into account the enormous cross-talk that has been identified between microbial and human cells (Blaser et al 2013) particularly in relation with the gut microbiome and the nervous and 
immune systems (Andrey Smith 2015). Engineering such microbial circuits 
is a major challenge that requires moving beyond the sense-and-deploy framework. 

Building complex decision making circuits within a single cell is a challenging task but 
several candidates have been suggested (Amos 2004; Fernando et al 2009; Lu et al 2009; Sorek et al 2013; 
Sardanyes et al 2015; Sol\'e et al 2015). These studies propose different ways 
of approaching the problem of building synthetic systems capable of diverse 
forms of bacterial intelligence. One example is the associative learning circuit presented 
by Fernando and co-workers that could be implemented in {\em E. coli} as a model 
organism  (Fernando et al 2009). It was inspired in previous theoretical studies that used model 
neural networks to explain the process under a minimal set of assumptions. 
In this case, the problem with the proposed design (as well as many others) is that it requires 
engineering several interactions, tuning the connectivity matrix of the molecular network, 
with all the problems derived from cross-talk (Kwok 2010). \\

In this paper we aim to provide a simple short cut to this problem, by using cellular consortia of 
cells that are used as basic modules, each one containing 
a small amount of engineering. This approximation has been successful in different contexts 
(Regot et al 2011; Tamsir et al, 2011; Macia and Sol\'e 2013, 2014, 
Goni-Moreno et al 2013; Macia et al 2016). 
In the next section, we describe the logic 
of our system design in order to illustrate its simplicity. Next, a potential implementation using 
a computational model for {\em E. coli} will be described.

\section{The logic of multicellular learning}

A synthetic circuit capable of associative learning requires some type of modulation of internal states 
through the learning process. Since the circuit responds to one signal (US) but not the other (CS) unless they have 
been previously presented together, this indicates that the internal states of the underlying molecular circuit 
must have changed. In figure 1a we show an example of a genetic implementation of AL introduced 
in (Sorek et al 2013). This work proposed a design inspired in neural networks. Here $X$ can activate the 
response $R$ whereas $Y$ will do it (when $X=0$) only if an intermediate module $M$ (that needs to be previously 
activated by $X+Y$) has the right expression level. This kind of design and others of similar inspiration 
(Fernando et al 2009) rely, if implemented inside one cell, a sophisticated wiring. 

\begin{figure*}[htp]
\begin{center}
\includegraphics[width=0.85 \textwidth]{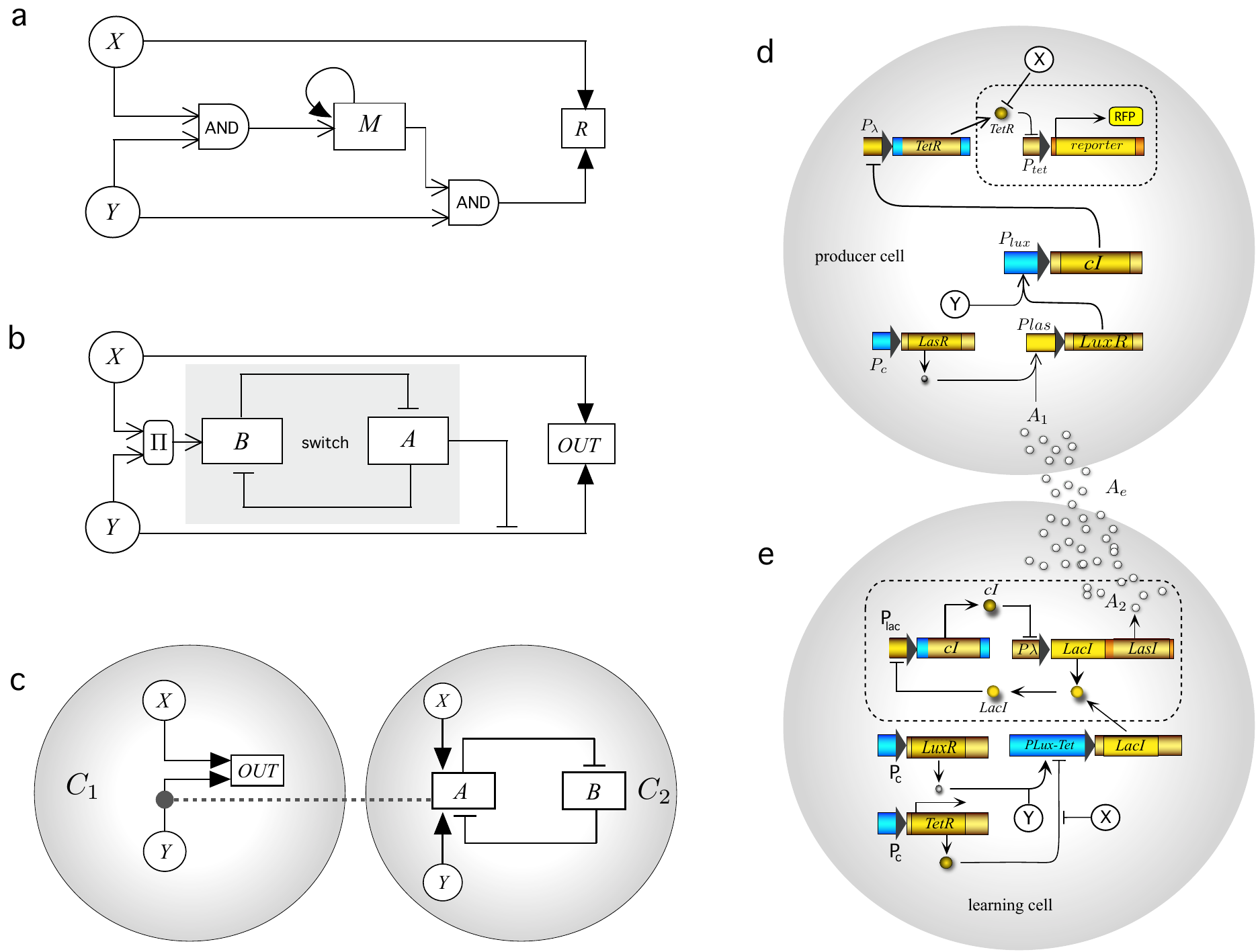}
\caption{The logic of a two-cell associative learning circuit. In (a) an example of a "standard" 
implementation, following neural network principles, is shown (redrawn from Sorek et al 2013). 
In (b) we summarize the basis of the circuit presented here, that exploits the presence 
of a toggle switch (indicated in gray). Two inputs are present, $X$ and $Y$ indicating unconditioned and 
conditioned signals, respectively. The circuit can be split in two parts corresponding to two engineered 
cells (c) here indicated as $C_1$ and $C_2$.  In (c-d) the proposed implementation of the synthetic 
consortium is shown. Here the two cells (d, producer, e, learning cell) communicate in one direction by 
means of a molecular signal ($A$).  Each engineered cell type performs 
part of the processing required to implement the association 
mechanism. We have used specific genes, cell-cell communication signals and reporters, 
but the basic principle can be used in different contexts (see text). }
\label{fig_model}
\end{center}
\end{figure*}

One of the most fundamental requirements for associative learning is memory, and designing 
synthetic circuits  (Gardner et al 2000; 
Ajo-Franlin et al  2010; Fritz et al 2007; Siuti et al 2013, Padirac et al 2012, 
Inniss and Silver 2013; Burrill and Silver 2010). 
A well known, successful example of memory circuit 
is provided by the toggle switch (Gardner et al 2000; Cherry and Adler 2000; Rodrigo and Jaramillo 2007)). 
This module has been extensively studied and characterised and 
is one of the best known components in cellular engineering. Since a molecular switch is capable 
of storing two alternative states, we use it here as a key piece of our proposed multicellular circuit. This introduces 
a restriction within the design of the system in relation to previous models. In order to illustrate 
how we perform our implementation, in figure 1b we represent a basic wiring diagram where two inputs 
are indicated as $X$ and $Y$, corresponding to the unconditioned and conditioned signals, respectively. 

\begin{table}[ttbp]
\begin{center}
\begin{tabular}{|cc|cc|c|}
\hline
X (Uncond) & Y (Cond) &  $\; \;$   State A  $\; \;$  & State B $\; \;$ &  Output \\ \hline
0 &  0 & 1 &  0 & 0 \\   \hline
0 &  1 & 1 &  0 & 0 \\  \hline
1 &  0 & 1 &  0 & 1 \\  \hline
1 &  1 & 0 &  1 & 1 \\  \hline
0 &  0 & 0 &  1 & 0 \\  \hline
0 &  1 & 0 &  1 & 1 \\  \hline
1 &  0 & 0 &  1 & 1 \\ \hline
\end{tabular}
\end{center}
\caption{Transition table for the simplified Boolean circuit implementing the two-cell circuit shown in figure 1c. The 
different input pair values given in the two left columns provides the sequence of states ($X$ and $Y$ 
for the conditional and unconditional inputs) introduced to test the 
presence of associative learning, assuming that $A=1$ and $B=0$ at the beginning.}
\end{table}

The diagram in figure 1b presents some similarities with the one in figure 1a. We will assume here that 
the switch has an internal, initial state, with $B=0$ and $A>0$. If $X>0$ and $Y=0$, a response will be observed 
since the response unit receives direct and positive stimulation from X. Instead, since $A$ inhibits 
the potential activation from $Y$, a signal coming inly from $Y$ will not trigger response. However, if 
both $X$ and $Y$ activate $X$, it can toggle the switch, inactivating (or under-expressing) $A$. Once this 
simultaneous activation occurs, the system is ready to react to $Y$ only. This defines the basic logic of our 
implementation, but we have split the circuit in two parts (figure 1c) corresponding to a {\em learning cell}, 
$C_1$, carrying the toggle switch and a producer cell $C_2$, that is wired to $C_1$ 
through a molecular wire $A$. As can be seen, we maintain 
the same scheme, but use cells as modules that allow to reduce the complexity of the engineering. 
In the next section, we make an explicit case for a microbial consortium capable of AL.

Because of the large number of equations involved, little mathematical developments can be performed and the solutions 
will be numerical. However, it is possible to see how the model works (and predict the key outcomes) 
by using a simple Boolean circuit as the one indicated in figure 1c. Here we use a discrete dynamical 
system based on a threshold network where all states are either $0$ or $1$. A reporter signal ($OUT$) with two possible states provides 
the result of the computation. In the middle of the circuit we have located a module involving cross-repression 
of two elements ($A$ and $B$) one of which can also modulate (through an inhibitory interaction) the effect of 
$Y$ on the output. We can actually represent these interactions in terms of a Boolean dynamical system, 
where the state of each element $S_i(t) \in \{ X, Y, A, B, O \}$ at a given step $t$ (assuming time is discretized) 
follows a discrete threshold dynamics: $S_i(t+1) = \Phi \left [ W_{ji} S_j(t) - \theta_i \right ]$. 
Here the connections among different pairs are indicated as $ W_{ji}$ and can be positive or negative, 
indicating activation or inhibition respectively (Grossberg 1988). In figure 1a 
we have included an example of possible weights for our system. The thresholds $\theta_i$ provide a condition 
for the total input in order to trigger a response. This is defined by the threshold function $\Theta(x)$ which gives 
$\Theta(x)=1$ if $x\ge 0$ and zero otherwise. This ideal function is a limit case of the standard cooperative functions 
used in models of genetic networks (see section III). 

For the circuit described in figure 1, our (discrete) equations are written as follows:
\begin{eqnarray}
A(t+1) = \Phi \left[ -B(t) + \theta \right] \\ 
B(t+1)= \Phi \left[ X(t) - Y(t) (1- A(t)) - \theta \right] \\ 
O(t+1) = \Phi \left[ X(t) + Y(t) \right]
\end{eqnarray}
It is possible to show, following the discrete steps of this Boolean model, that an associative 
learning dynamics is being satisfied. The sequence of states associate to the consortium displayed 
in fig 1c is shown in Table I, where the set of possible input pairs $(X,Y)$ and the $(A,B)$ the states 
of the elements defining the memory switch (SM).

\section{Associative learning in a synthetic microbial consortium}

In order to avoid undesirable effects derived from complex constructs, cellular consortia, where 
different parts of the computation are split into different engineered cells, can be used as an 
alternative to single-cell designs. An example of such synthetic consortium  is displayed in figure 2(d-e). 
It combines both constitutive and regulated gene expression and splits the circuit complexity 
in two separated cells. As summarised in figurers 1b-e, the required behaviour is split into two basic modules, each one using a different engineering. Although we will assume that the two cell types belong to the same species, this is not a necessary condition. Each cell in this consortia acts as a separated  chassis for a subset of the required circuit. In our proposed implementation, we will use available information concerning 
well established constructs and parameters gathered from the available literature 
and take {\em E. coli} as our model organism. Several potential candidates could be used as 
inputs, such anhydrotetracycline (aTc) as our non-conditional stimulus ($X$) and 
Acyl homoserine lactone (AHL) (produced the gene luxI  from the {\em V. fischeri} quorum sensing system) 
as a conditioned stimulus ($Y$). In presence of this stimulus there is not response unless the system 
has established the association between $X$ and $Y$. In a nutshell, 
whereas $X$ alone always induces a system's response, $Y$ does not. The computational model 
requires an explicit definition of the mathematical equations for each cell, as well as 
the consideration of the 

We have chosen a fluorescent protein as the candidate for the cell's output, although this 
could be some gene that triggers the delivery of a given therapeutic molecule. 
LuxR is a transcriptional activator from the {\em V. fischeri} quorum sensing system
that binds to its cognate promoter Plux activating the expression of genes under its control. The wild type LuxR
is inactive when produced. Acyl homoserine lactone (AHL), produced by another gene, luxI, is an autoinducer
that binds LuxR and increases its activity. 

\begin{figure*}[t]
\begin{center}
\includegraphics[width=0.85 \textwidth]{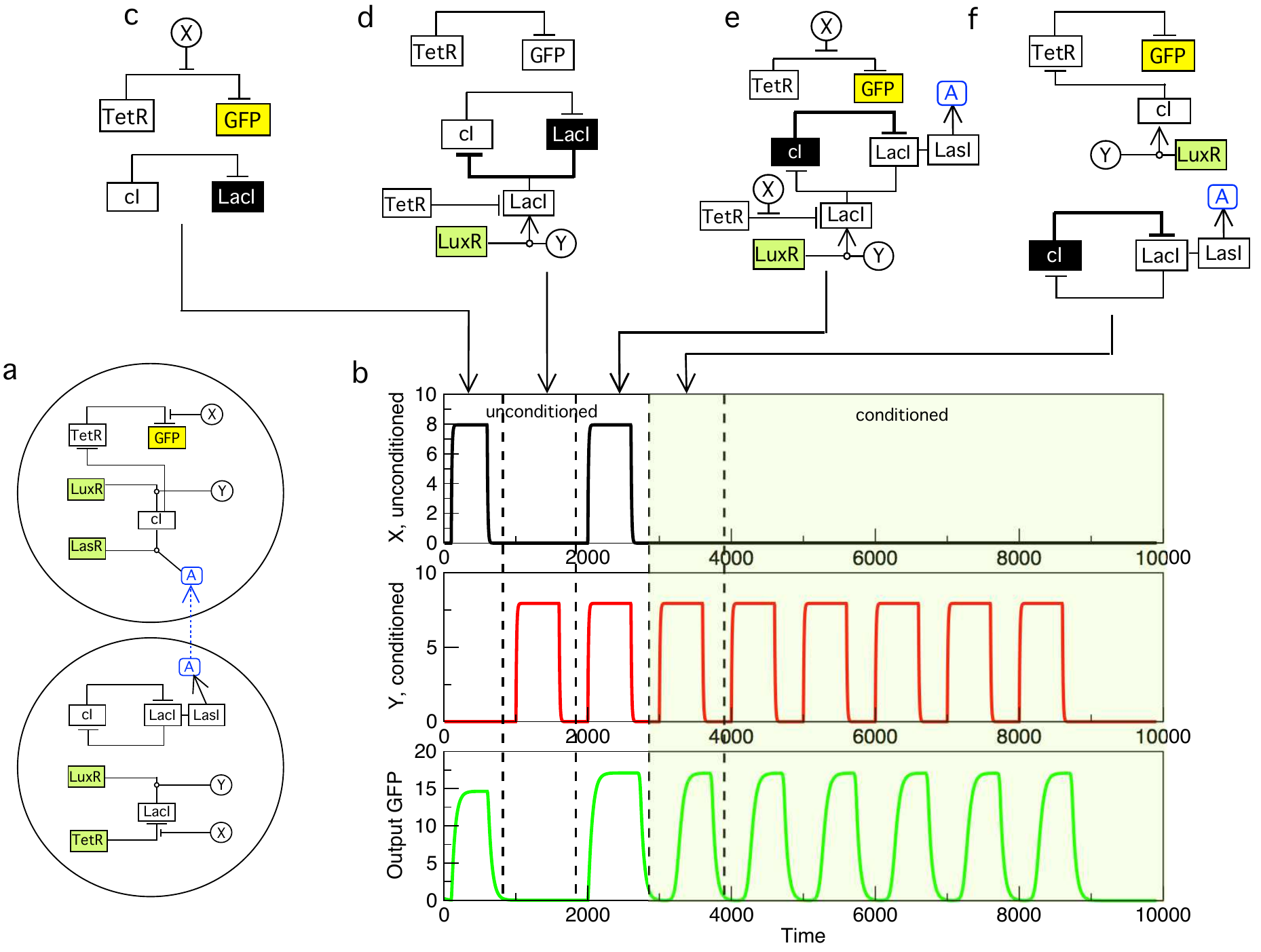}
\caption{Time series of the circuit behavior under the sequential introduction of the unconditional stimulus $X$ the two types of stimuli. In (a) we show the basic diagram of interactions detailed in figure 2. In figure b we show the time 
series associated to an experiment of classical conditioning at the cell level. The first two time series give the 
inputs of $X$ and $Y$, introduced as pulses for a given period and then removed. The sequence involves 
$X>0, Y=0$, $X=0, Y>0$, $X,Y>0$ and then several pulses of $Y$-only activation. The upper diagrams (c-f) 
represent the relevant subgraphs that are changed or activated once the pulse has been introduced and the system stabilised. Here we have used the parameters: 
$\gamma_{TetR}=\gamma_{Las}=\gamma_{cI}=\gamma_{LuxR}=\gamma_{LacI}=\gamma_{A}=\gamma_{\lambda}=\gamma_{GFP}=0.1 \; \mu M/min$, $\beta_{cI}=0.008  \; \mu M$, 
$\beta_{Tet}=0.04  \; \mu M$, $\theta_{Lux}=\theta_{Las}=0.01  \; \mu M^2$, $\delta_{TetR}=\delta_{Las}=0.02  \; min^{-1}$, 
$\delta_{cI}=0.07  \; min^{-1}$, $\delta_{LuxR}=\delta_{GFP}=0.02  \; min^{-1}$, $\mu=0.025  \; \mu M$, $D_A=0.1 \; min^{-1}$. 
}
\label{AssocLearningSeries}
\end{center}
\end{figure*}

The mathematical model associated to the cellular consortium displayed in figure 2 is decomposed in 
two sets of equations. Both cells have $X$ and $Y$ as inputs, but the nature of their responses 
is markedly different.

\subsection{Producer cell equations}

For the producer cell, we have five coupled differential equations, describing the 
basic interactions indicated in figure 1d. These equations are standard 
in the modelling of gene regulation networks (Ingalls 2103). Here we have 
a feed-forward set of interactions described by:
\begin{equation}
{d [LasR] \over dt} = \gamma_{LasR}  - \delta_{Las} [LasR]
\end{equation}
which is a constitutive gene (here $P_c$ will indicate a constitutive promoter). The dynamical 
equations for the rest of components in our cellular circuit read:
\begin{equation}
{d [LuxR] \over dt} = \gamma_{Lux}\Gamma_2([LasR],[A]) - \delta_{LuxR} [LuxR]
\end{equation}
\begin{equation}
{d [cI] \over dt} = \gamma_{cI} \Gamma_1([LuxR],[Y]) - \delta_{cI} [cI]
\end{equation}
\begin{equation}
{d [TetR] \over dt} = {\gamma_{TetR} \over 1 + \left (  {[cI] \over \beta_{cI}}  \right )^2 } - \delta_{TetR} [TetR]
\end{equation}
And the equation for the response dynamics, described by the concentration of our reporter, 
is defined by: 
\begin{equation}
{d [GFP] \over dt} = \gamma_{GFP} \Gamma_3([TetR],[X]) - \delta_{GFP} [GFP]
\end{equation}
The Hill functions used here are described by the functions 
$\Gamma_1 ([LuxR],[Y])$ and $\Gamma_2([LasR],[A])$ involving 
thresholded activation:
\begin{equation}
\Gamma_1([LuxR],[Y]) = { \left([LuxR][Y]\right)^2 \over \theta_{Lux} + \left([LuxR][Y]\right)^2} 
\end{equation}
\begin{equation}
\Gamma_2([LasR],[A]) =  { ([LasR][A])^2 \over \theta_{Las} + \left([LasR][A]\right)^2} 
\end{equation}
along with the Hill inhibition function:
\begin{equation}
\Gamma_3([TetR],[X]) =  {1 \over 1 +  \left ( {[TetR] \over \beta_{Tet}}� (1 +[ X]/\mu) \right )^2 } 
\end{equation}
In particular, we can see that the reporter will be active if no repression from TetR is at work. 
Either by inactivation of TetR or by the presence of $X$, the response will be observed.

\begin{figure}[!]
\begin{center}
\includegraphics[width=0.43 \textwidth]{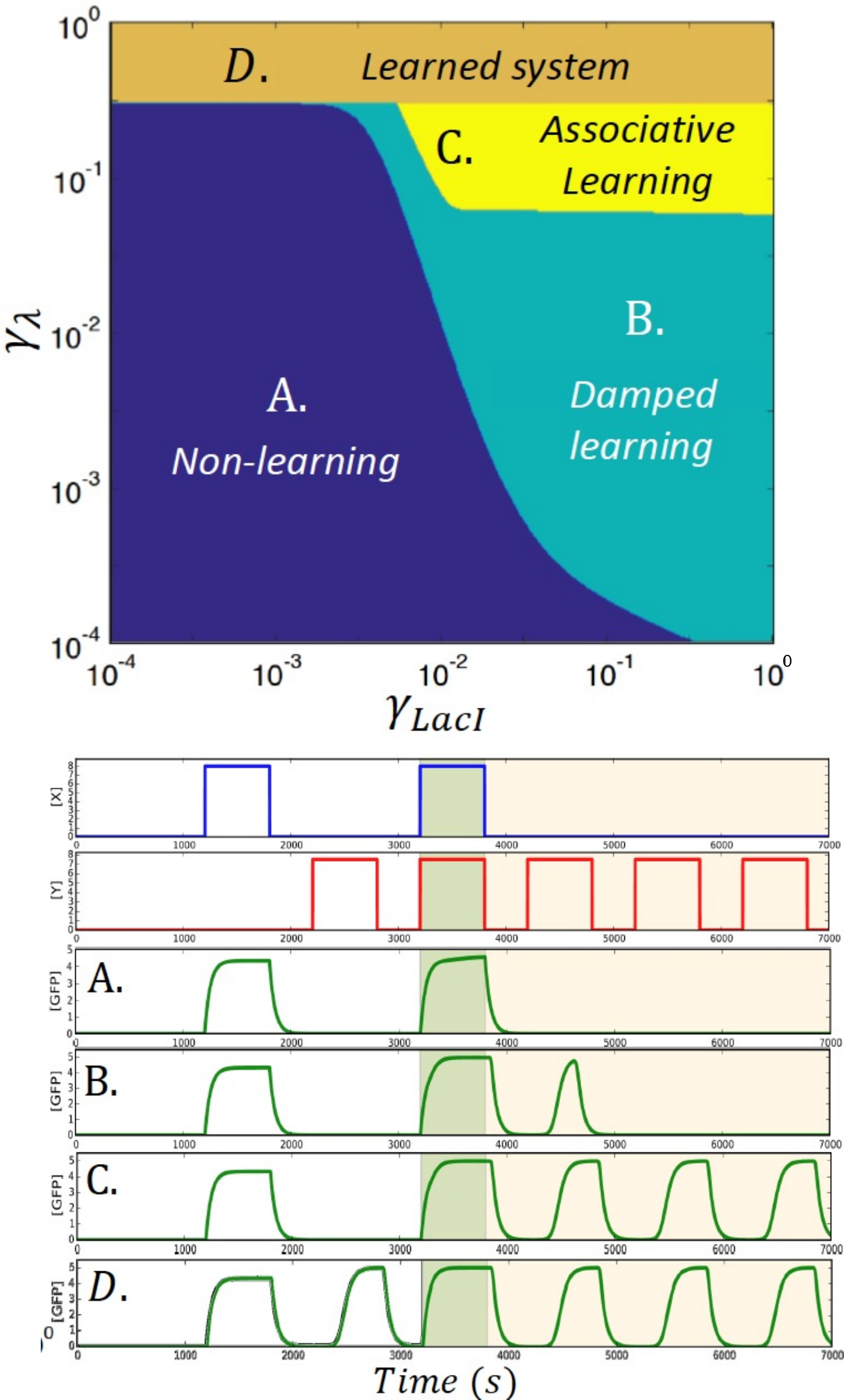}
\caption{Phase space of the associative learning two-cell design and their output time series. Depending on the production rate of LacI ($\gamma_{LacI}$) and the production rate of LacI under CI regulation ($\gamma_\lambda$) there are four different response behaviours. (A.) Non learning situation; the system responds to the unconditioned stimuli ($X$) (blue region), the system is unable to respond to the conditioned signal ($Y$) even after the conditioning stimuli process.(B.) Damped learning; the learning cell has a temporal association (turquoise), the producer cells only respond to the conditioned signal for a limited time after the conditioning. (C.) Associative learning region (yellow), the learning cell change the state of the toggle switch after the simultaneous stimulation with $X$ and $Y$ signals, then the system have learned. (D.) Learned system (orange); the system responds to the unconditioned ($X$) and conditioned ($Y$) signals independently before the conditioning process pulses. The time series of the other variables can be seen at the supplementary figures $3$-$6$.   The parameters are the same than in previous figures; the phase space is performed with a $400x400$ latice. 
 }
\label{fig_behaviours}%
\end{center}
\end{figure}

\subsection{Learning cell equations}

 For the learning cell (figure 1e) we consider a different set of equations. Here two 
 genes are expressed constitutively thus involving linear equations: 
\begin{equation}
{d [LuxR] \over dt} = \gamma_{LuxR}  - \delta_{LuxR} [LuxR]
\end{equation}
\begin{equation}
{d [TetR] \over dt} = \gamma_{TetR} - \delta_{TetR} [TetR]
\end{equation}
which provide the gene products that will interact with $X$ and $Y$ within this cell under 
the nonlinear equtions
\begin{equation}
{d [LacI] \over dt} = \gamma_{LacI} \Gamma_3 \Gamma_1
+ {\gamma_{\lambda} \over 1 + \left (  {[cI] \over \beta_{cI}}  \right )^2 } 
- \delta_{LacI} [LacI]
\end{equation}
We have also the well-known equations for the toggle switch defined by the pair: 
\begin{equation}
{d [cI] \over dt} =  {\gamma_{cI} \over 1 + \left (  {[LacI] \over \beta_{Lac}}  \right )^2 }  - \delta_{cI} [cI]
\end{equation}
\begin{equation}
{d [LasI] \over dt} = {\gamma_{\lambda} \over 1 + \left (  {[cI] \over \beta_{cI}}  \right )^2 } - \delta_{LasI} [LasI]
\end{equation}
Which have two alternative states. Finally the linear equation 
for the production of the molecule $A$, used in our first model 
as the communication signal among the two cells:
\begin{equation}
{d [A] \over dt} = \gamma_{A} LasI  - \delta_A [A]
\end{equation}

\subsection{Cell-cell communication wire}

A final component needs also to be taken into account: the diffusion of the wiring molecule $A$
responsible for the intercellular connection. The last equation above only considers the production 
within $C_2$, but it is shared with cell $C_1$ by diffusion and is also present in the 
extracellular medium ($A_e$). Thus we need to write three equations that account for the dynamics of $A$ in each 
compartment (see figure 2). The complete equations read: 
\begin{eqnarray}
{d [A_2] \over dt} = \gamma_{A_2} LasI  - \delta_A [A_2] + D_A ([A_e]  - [A_2]) \\
{d [A_1] \over dt} =  D_A ([A_e]  - [A_1])- \delta_A [A_1]\\
 {d [A_e] \over dt} =  D_A ([A_1]+[A_2]-2[A_e])- \delta_A [A_e]
\end{eqnarray}
corresponding to the two cells and the extracellular medium, respectively. 
 
\subsection{Associative learning dynamics}

The previous equations start from an initial condition where the toggle switch is displaced 
towards $cI$. This defines the memory state of our system at time zero. Each input is introduced in the system in a pulse-like way. In figure 2 we show a typical example 
of the numerical experiment consistent with an associative learning process. The left diagram 
(figure 2a) provides a schematic representation of all the interactions and figure b 
shows the time series obtained from the model. We first start by introducing $X$ 
but not $Y$. This leads to a response as shown by the pulse in GFP, which disappears as $X$ is also removed from the system. The positive response is easy to understand, since the only pathway being affected leading to GFP 
is indicated in figure 2c, where$X$ blocks the inhibition of the reporter from $TetR$. 

Afterwards, we do the same experiment with $Y$ but in this case no active reporter is seen. The repressor of GFP 
acts with no inhibition (figure 2d) and the paths affected by $Y$ do not propagate. The crucial change occurs when the two inputs are simultaneously correlated. Here the reporter is again activated (top of diagram e) as it 
happened in the first $X$-only pulse. However, the effect of the simultaneous input on the toggle switch 
is that the state of the cI-LacI pair switches to the opposite state, where CI is now expressed and 
LacI repressed. This is the internal state that has changed as a consequence of the correlated stimulus 
and will remain in this state once we remove both inputs. 

Once the previous pulse has been applied and both inputs removed again, we can see the effect of 
these correlated input when the conditioned, $Y$ signal is introduced in the absence of the unconditioned one. 
Here the stored memory state in the toggle switch has a very different impact. In this case, this state is 
not changed, but allows the propagation of the effects of $Y$ to the producer cell, where cI is produced, repressing TetR and thus allowing GFP to be expressed. The consortium has created an association (thanks to the toggle) that essentially modifies the system's response to the conditioned state. 

The model presented above has been analysed using a given parameter combination. 
What is the effect of other parameter values on the dynamical response of 
the consortium to other parameter combinations. Two parameters in particular 
are relevant to our exploration of the response of the system. These are the production 
rates $\gamma_{Lac}$ and $\gamma_{\lambda}$. By exploring the 
$(\gamma_{Lac},\gamma_{\lambda})$ parameter space (fig 3a) using a wide range of parameter 
values i. e. $10^{-4} \le \gamma_{Lac},\gamma_{\lambda} \le 10^0$. Four dynamical phases 
are found, which are associated to four different types of responses to the incoming stimuli.  

As in previous sections, the synthetic consortium receives a sequence of inputs where the 
unconditioned stimulus $X$ is used first, followed by the conditional one $Y$ and then both 
together. Afterwards, pulses of $Y$ are introduced and the type of output response is used 
to classify the circuit's behaviour. The right panel in figure 3 shows examples of the output 
response for each phase as the sequence of stimuli is introduced. The four classes are captured 
by the time series associated to each one (A-D) in figure 3 right. No learning occurs within one phase 
where there is only unconditional response. In a second phase (Damped learning, B, see figure 3 left) 
a small responses is observed suggesting association, but it is rapidly lost after one weak response 
to the conditional stimulus. A yellow domain points to the associative learning parameter space 
whereas the domain of learned systems is associated to responses by both stimuli, no matter 
how are they presented. The plot has been created on a log-log 
scale, and thus we can see that a broad range of parameters are consistent with this behavior.

\subsection{Supresing associative learning}

\begin{figure*}[!]
\begin{center}
\includegraphics[width=0.9 \textwidth]{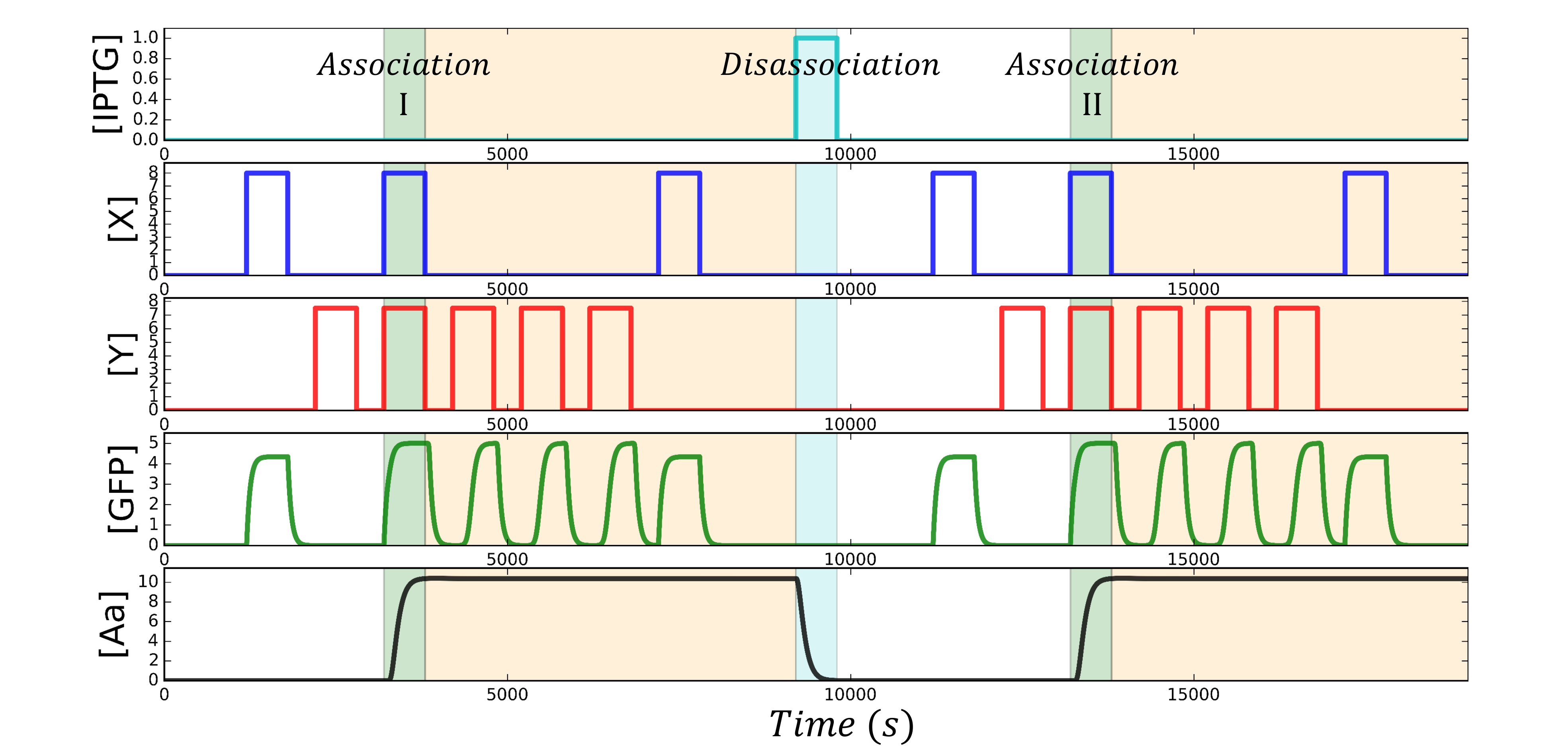}
\caption{Time series of conditioning, memory erase and conditioning again. White regions are the non-conditioned regions; the cells are onlly able to express the GFP molecule under $X$ signalling. The green regions are regions where the system is being conditioned, there is the association process, simultaneous stimulation with $X$ and $Y$. Orange regions the system is conditioned; it is able to responds to the unconditioned stimuli and to the conditioned. When there is a pulse of IPTG (cyan area) there is a loss of memory, a disassociation between signals. The parameters are the same than in previous figure and $\beta_{IPTG}=0.04$ and the $IPTG$ pulse is $1  \; \mu M$ of amplitude.
 }
\label{fig_Association_Disassociation}
\end{center}
\end{figure*}

Once the association has been established in our circuit, as it occurs with conditioned 
learning in animals, the switch is locked in a given state that allows the association 
to be stable over time. However, if this is a circuit that has been designed to respond 
to unconditional stimuli once the inputs of both kinds are presented simultaneously. 
It can be interesting to return to the initial state where the consortium has not 
yet

The LacI protein have one well known inhibitor, the IPTG molecule, which 
\begin{equation}
\frac{d [CI]}{dt} = {
\gamma_{cI} 
\over 
1 +  \left (  {[LacI]  \over \beta_{Lac}  \left ( 1+ \xi \right)} \right )^2 
}
- \delta_{cI} [cI] 
\end{equation}
where $\xi=IPTG/\beta_{IPTG}$. A pulse of IPTG 
makes inhibition of the LacI function. Then, the 
CI promoter is active again leading to an inversion of the toggle 
switch LacI-CI. After this process the conditioning have been lost.

\section{Discussion}

\begin{figure*}[!]
\begin{center}
\includegraphics[width=0.9 \textwidth]{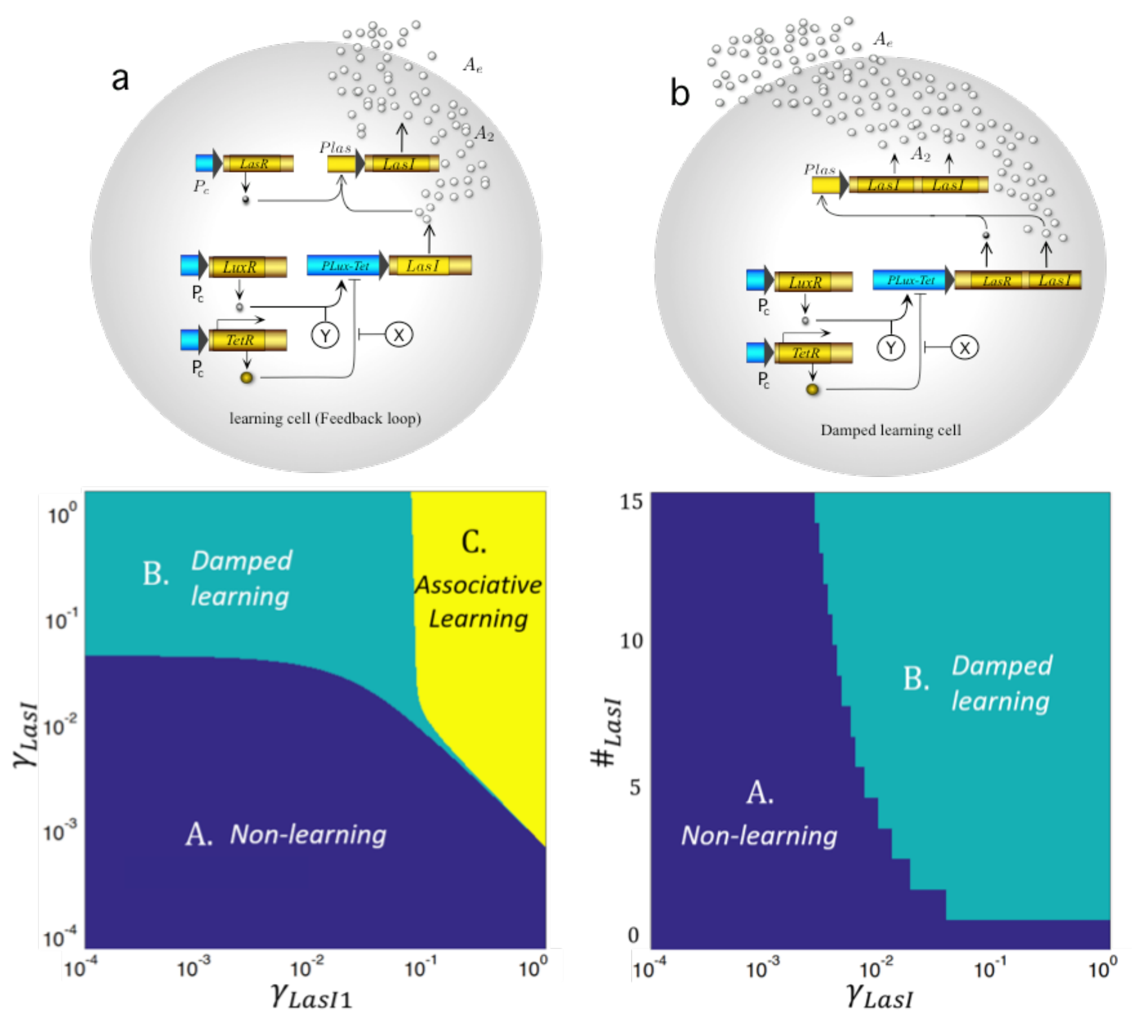}
\caption{Alternative circuits for synthetic memory. a. Associative learning circuit based on positive feedback loop (FIG.1a). b. Damped learning synthetic circuit. Bellow each diagram, there is the phase space relating the parameters with the behaviour exhibited: (A.) There is response to the unconditioned signal. (B.) Damped learning, there is a characteristic time where the two cells system responds to $X$ and $Y$, afterwards  the system does not respond to the unconditioned signal. (C.) Associative learning, once the cells are stimulated simultaneously by both molecules ($X$ and $Y$), the system responds either the conditioned or unconditioned stimulus. The mathematical models are exposed in the Supplementary Material. The parameters are set to the same values than in FIG.2. The phase space are lattices of $400$ units per parameter, excepting the number of copies of LasI given that this parameter is discrete. 
 }
\label{fig_AL_alt}
\end{center}
\end{figure*}

One of the challenges synthetic biology is to make possible the
reprogramming of cellular behaviour by means of a predictable, engineered manipulation of 
the available molecular toolkit. The potential of such engineered molecular networks is great, 
and cover a wide range of areas, from standard biosensors to complex decision making circuits 
able to gather a range of external stimuli from the environment and respond according to a 
predefined set of rules.  An important goal is to provide these engineered systems with 
the appropriate adaptation potential, which necessarily requires the use of learning and memory. 
In this context, the potential for recapitulating the evolutionary innovations by building synthetic circuits 
provides a unique opportunity for the study of major transitions (Sol\'e 2016). 

The proposed synthetic systems presented here show that the use of a cell consortia 
can help designing complex decision-making biological circuits capable to cope with external 
signals and their changes. The human microbiome provides an ideal testbed for these kind of 
synthetic designs. If the metaphor of this as a "second brain" becomes valid, then what we are 
suggesting is to introduce pieces of computational complexity to play an active role within 
the network of microbial interactions. Our example also combines the internal machinery that 
responds to external signals  (which could be drugs) with a flexible design capable 
of exploiting the history of previous events. This basic scheme can be generalised to more complex 
designs. Future work should test the experimental feasibility of our approach as well its scalability. 

The circuits proposed above assume that the two-cell consortium is obtained by engineering 
the same class of model organism, but this is not required for our purposes. Mixed consortia 
involving both microbial and human cells could be constructed, and other possibilities are also 
available, including the design of symbiotic consortia between soil microorganisms and plant cells 
(or nitrogen-fixing microorganisms living within plant nodules) as potential strategies of ecosystem 
repair (Sol\'e 2015, Sol\'e et al 2015). Concerning diseases associated with a malfunctioning microbiome, 
our results suggest that both permanent and transient modifications of some engineered 
strains could help to dynamically control some key processes requiring memory and learning. This is 
an interesting possibility given the feedback existing between both the immune system and the brain 
as connected with the microbiome (Mayer et al 2014; Sampson and Mazmanian 2015). 
Since both immune and brain networks are capable of displaying 
learning and memory, microbial consortia as the ones presented here 
could act as extensions of neural-like decision circuits. 

Finally, another interesting possibility concerns the 
design of synthetic learning circuits that could be incorporated within organoids (Lancaster and Knoblich 2014). 
Such enhanced cognitive complexity seems a desirable trait of future designed organoids and allow 
to move away from natural designs, thus reaching some unexplored regions of organ space (Oll\'e-Vila et al 2016).


\section{Acknowledgements}
We thank the members of the Complex Systems Lab for useful discussions. 
This work has been supported by an ERC Advanced Grant Number 294294 
from the EU seventh framework program (SYNCOM) the Botin Foundation by 
Banco Santander through its Santander Universities Global Division, the Secretaria 
d'Universitats i Recerca del Departament d'Economia i Coneixement de la Generalitat de 
Catalunya, a grant of MINECO, FIS2015-67616-P and by the Santa Fe Institute.


\bibliography{pre}


\section{References}

\begin{enumerate}

\item
 Ackerman J (2012) The ultimate social network. Sci. Am.  June, 36-43.

\item
Ajo-Franklin, C. M., Drubin, D. A., Eskin, J. A., Gee, E. P. S., Landgraf, D., Phillips, I., and Silver, P. (2007). Rational design of memory in eukaryotic cells. Genes Dev. 21, 2271-2276.

\item
Amos, M. 2004. Cellular Computing. Oxford University Press, New York.

\item
Auslander, S, Ausl\"ander D, Muller M, Wieland M. and Fussenegger M.. 2012. 
Programmable single-cell mammalian biocomputers. Nature 487, 123-127.

 \item
Basu, S., Gerchman, Y., Collins, C. H., Arnold, F. H. and Weiss, R. 2005. A synthetic multicellular system for programmed pattern formation. Nature 434, 1130-1134. 

\item
Ben-Jacob, E. 2009. Learning from bacteria about natural information processing. 
Ann. N. Y. Acad. Sci. 1178, 78-90.
 
\item
Benenson, Y. 2012.   Biomolecular computing systems: principles, progress and potential. Nature Rev. Genet. 13, 455-468.

\item
Blaser M, Bork, P, Fraser C, Knigth R. and Wang J. (2013) The microbiome explored: recent insights and future challenges.  Nat. Rev. Microbiol. 11, 213-217 .

\item
Bray D (1995) Protein molecules as computational elements in living cells. Nature 376, 307-312.

\item
Bray D (2003) Molecular networks: the top-down view. Science 26: 1864-1865.

\item
Buchler NE, Gerland U, Hwa T (2003) On schemes of combinatorial transcription logic. 
Proc Natl Acad Sci U S A 100: 5136-5141.

\item
Burrill, D.R., and Silver, P. A. (2010). Making cellular memories. Cell 140, 13-18.

\item
Cherry, J. L. and Adler, F. R. 2000 How to make a biological switch. J. Theor. Biol.  203, 117-133.

\item
Fernando, C. T., Liekens, A. M., Bingle, L. E., Beck, C., Lenser, T., Stekel, D. J., and Rowe, J. E.  (2009). Molecular circuits for associative learning in single-celled organisms. J. Roy. Soc. Interface 6, 463-469.

\item
Friedland, A. E. et al. 2009. Synthetic gene networks that count. Science 324, 1199-1202

\item
Fritz, G., Buchler, N. E., Hwa, T., and Gerland, U. 2007. Designing sequential transcription logic: a simple 
genetic circuit for conditional memory. Syst. Synth. Biol. 1, 89-98.

\item
Gardner, T. S., Cantor, C. R., and Collins, J. J. (2000). Construction of a genetic toggle switch in \textit{Escherichia coli}. Nature 403, 339-342.

\item
Gerstner, W. and Kistler, W. M. (2002). Mathematical formulations of Hebbian learning. Biological cybernetics, 87(5-6), 404-415.

\item
Ginsburg, S. and Jablonka, E. (2010). The evolution of associative learning: A factor in the Cambrian explosion. J. Theor. 
Biol. 266, 11-20.

\item
Goni-Moreno, A. and Amos, M. 2013. Continuous computation in engineered gene circuits. 
Biosystems 109, 52-56

\item
Grossberg, S. 1988. {\em Neural networks and natural intelligence}. MIT Press, Cambridge MA. 

\item
Hassoun M. H. (ed) 1993.  {\em Associative neural networks. Theory and implementation}. Oxford U. Press. New York. 

\item
Hennessey TM, Rucker WB, Mcdiarmid CG. 1979. Classical-Conditioning in Paramecia. Anim. Learning  Behav 7: 417-423.

\item
Huttenhower C et al 2012 Structure, function and diversity of the healthy human microbiome. Nature 486: 207-214.

\item
Inniss, M.C. and Silver, P. A. 2013. Building Synthetic Memory. Current Biol. 23, R812-R816. 

\item
Kwok, R. 2010. Five hard truths for synthetic biology. Nature 463, 288-290.

\item
Lancaster, M. A. and Knoblich, J. A. 2014. 
Organogenesis in a dish: modeling development and disease using organoid technologies. 
Science 345, 1247125.

\item
Li, B. and You, L. 2011. Synthetic biology: Division of logic labour. Nature 469, 171-172.

\item
Lu, T., Khalil, A. S. and Collins, J. J. 2009. Next-generation synthetic gene networks. 
Nature Biotech. 27, 1139-1150.

\item
Macia, J., Posas, F., and Sol\'e, R.V. 2002. Distributed computation: the new wave of 
synthetic biology devices. Trends in Biotech. 30, 342-349. 

\item
Macia, J. and Sol\'e, R. V. 2014. 
How to Make a Synthetic Multicellular Computer. PLoS ONE 9, e81248.

\item
Macia, J., Manzoni, R., Conde, N., Urrios, A., de Nadal, E., Sol\'e, R. and Posas, F. 2016. 
Implementation of complex biological logic circuits using spatially distributed multicellular consortia. 
PLoS Comput. Biol. 12, e1004685.

\item
Mayer, E. A., Knight, R., Mazmanian, S. K., Cryan, J. F. and Tillisch, K. 2014. 
Gut microbes and the brain: paradigm shift in neuroscience. J. Neurosci. 34, 15490-15496.

\item
Mitchell A, Romano GH, Groisman B, Yona A, Dekel E, et al. 2009. Adaptive
prediction of environmental changes by microorganisms. Nature 460, 220-286.

\item
Oll\'e-Vila, A., Duran-Nebreda, S., Conde-Pueyo, N., Monta\~nez, R. and Sol\'e, R. 2016. 
A morphospace for synthetic organs and organoids: the possible and the actual. Integrative Biol. 8, 485-503.

\item
Padirac, A., Fufii, T. and Rondelez, Y. 2012. Bottom-up construction of in vitro switchable memories. 
Proc. Nat. Acad. Sci. USA 109, E3212-E3220.
 
\item
Pepper JW and Rosenfeld S. 2012. The emerging medical ecology of 
the human gut microbiome. Trends Ecol Evol 27: 381-384.

\item
Purnick, P. E. M., and Weiss, R. 2009. The second wave of synthetic biology: 
from modules to systems. Nature Rev. Cell. Biol. 10, 410-422.

\item 
Reid, C. R. et al. 2015. Information integration. Anim. Behav. 100: 44-50.

\item
Regot, S., Macia, J., Conde, N., Furukawa, K. et al. 2011. 
Distributed biological computation with multicellular engineered networks. Nature 469, 207-211. 

\item
Rodrigo, G. and Jaramillo, A. 2007. Computational design of digital and memory biological devices. 
Syst. Synth. Biol. 1, 183-195.

\item
Sampson, T. R. and Mazmanian, S. K. 2015. Control of brain development, function, 
and behavior by the microbiome. Cell host $\&$ microbe 17, 565-576.

\item
Siuti, P., Yazbek, J., and Lu, T. K. 2013. Synthetic circuits integrating logic and memory in living cells. 
Nature Biotech. 31, 448-453. 

\item
Sol\'e R. et al 2015. Synthetic collective intelligence. Biosystems (to appear).

\item
Sol\'e, R.V. and Macia, J. 2013. Expanding the landscape of biological computation with synthetic multicellular consortia. \textit{Natural Comput.} 12, 485-497.

\item
 Sol\'e, R. 2015. Bioengineering the biosphere?. Ecological Complexity, 22, 40-49.

\item
Sol\'e, S., Monta\~nez, R., and Duran Nebreda, S. 2015. Synthetic circuit designs for earth terraformation. Biology Direct. 10: 37.

\item
Sol\'e, S. (2016) Synthetic transitions: towards a new synthesis. Phil. Trans. R. Soc. B, 371, 20150438.

\item
Sorek, M., Balaban, N. Q., and Loewenstein, Y. (2013). 
Stochasticity, Bistability and the Wisdom of Crowds: A Model for 
Associative Learning in Genetic Regulatory Networks. PLOS Comp. Biol. 9, e1003179.

\item
Tagkopoulos I, Liu YC, Tavazoie S. 2008. Predictive behavior within microbial
genetic networks. Science 320: 1313-1317.

\item
Tamsir, A., Tabor, JJ., and Voigt, C. A. 2011. Robust multicellular computing 
using genetically encoded NOR gates and chemical 'wires'. Nature 469, 212-215.

\item
Urrios, A., Macia, J., Manzoni, R., Conde, N., et al, R. 
2016. A synthetic multicellular memory device. ACS Synthetic Biol. , 862-873.	

\item
Walters, E. T., Carew, T. J. and Kandel, E. R. 1979 Classical conditioning in 
Aplysia californica. Proc. Natl Acad. Sci. USA 76, 6675-6679.

\item
Weber, W. and Fussenegger, M. 2012. 
Emerging biomedical applications of synthetic biology. Nature Rev. Gen. 13, 21-35.

\end{enumerate}

\end{document}